\documentclass{svproc}
\usepackage{url}

\usepackage{graphicx}
\usepackage{common}
\usepackage{flexisym}
\usepackage{multirow}
\usepackage{cite}

\usepackage{xcolor}
\usepackage{environ}
\usepackage{subfig}
\usepackage{tabularx}
\usepackage{array}
\usepackage{color}
\usepackage{wasysym}
\usepackage{multicol}
\usepackage[bottom]{footmisc}
\usepackage{amsmath}
\usepackage{lscape}

\usepackage{float}
\usepackage[ruled,vlined]{algorithm2e}
\usepackage[font=small,skip=3pt]{caption}
\newcolumntype{?}{!{\vrule width 1pt}}
\makeatletter
\newcommand{\thickhline}{%
    \noalign {\ifnum 0=`}\fi \hrule height 1pt
    \futurelet \reserved@a \@xhline
}
\newcolumntype{"}{@{\hskip\tabcolsep\vrule width 2pt\hskip\tabcolsep}}
\makeatother
\newcolumntype{M}[1]{>{\centering\arraybackslash}m{#1}}
\newcolumntype{N}{@{}m{0pt}@{}}

\begin{document}
\mainmatter              
\title{Semi-Supervised Overlapping Community Finding based on Label Propagation with Pairwise Constraints}
\titlerunning{Semi-Supervised Overlapping Community Finding}  
%
\author{Elham Alghamdi\inst{1} \and Derek Greene\inst{2}}
\authorrunning{Elham Alghamdi et al.} 
%
\tocauthor{Elham Alghamdi and Derek Greene}
\institute{University College Dublin, Ireland\\
\email{elham.alghamdi@ucdconnect.ie}
\and
University College Dublin, Ireland\\
\email{derek.greene@ucd.ie}}

\maketitle              

\begin{abstract}
Algorithms for detecting communities in complex networks are generally unsupervised, relying solely on the structure of the network. However, these methods can often fail to uncover meaningful groupings that reflect the underlying communities in the data, particularly when those structures are highly overlapping. One way to improve the usefulness of these algorithms is by incorporating additional background information, which can be used as a source of constraints to direct the community detection process. In this work, we explore the potential of semi-supervised strategies to improve algorithms for finding overlapping communities in networks. Specifically, we propose a new method, based on label propagation, for finding communities using a limited number of pairwise constraints. Evaluations on synthetic and real-world datasets demonstrate the potential of this approach for uncovering meaningful community structures in cases where each node can potentially belong to more than one community.
\keywords{overlapping community finding, semi-supervised learning}
\end{abstract}
%

\section{Introduction}
In many real-world application involving machine learning, the tasks do not neatly correspond to the standard distinction between supervised and unsupervised learning. Rather, a limited degree of background knowledge or user annotation time will be available. Tasks such as community detection can potentially benefit from the introduction of ``lightweight'' supervision originating from domain experts or crowdsourced annotations, where this knowledge might be encoded as constraints indicating that a pair of nodes should always be assigned to the same community or should never be assigned to the same community. For instance, we might be interested in grouping users on a social media platform such as Twitter, based primarily on their follower connections, in order to discover communities of individuals with shared ideologies. To improve our ability to achieve this, and go beyond simply looking at connections, we could present pairs of user profiles to a human annotator (the ``oracle''), to ask whether two users should be assigned to the same community or different communities. By harnessing this kind of knowledge, we can potentially uncover communities of nodes which are difficult to identify with unsupervised methods.

Initial work in community detection focused on the development of algorithms to produce disjoint groups \cite{blondel08fast}. However, in many real-world networks we observe pervasive overlap, where nodes belong to many highly-overlapping groups \cite{ahn10link}. More recently, overlapping community finding algorithms have been developed for application to these networks \cite{ahn10link,lee10gce}. However, this work has focused only on the unsupervised case. In contrast, work on semi-supervised community finding continues to focus on cases where communities are strictly required to be disjoint \cite{li2014extremal}.

In this paper, we propose a semi-supervised method for overlapping community finding based on a label propagation strategy, which has previously been applied in a purely unsupervised context \cite{xie11slpa}. The proposed method, referred to as Pairwise Constrained SLPA (PC-SLPA), involves a speaker-listener information propagation process. To encode external supervision, we use pairwise constraints to influence the community finding process. Since the choice of constraints in semi-supervised learning has been shown to be highly important \cite{leng2013active}, we further propose a strategy for selecting constraint pairs for which an oracle should be queried. This strategy is specifically designed for the case where communities overlap in a network. The experiments described later in Section \ref{sec:eval}, which involve synthetic and real networks, show that the introduction of a relatively small number of constraints with PC-SLPA can improve our ability to correctly uncover the underlying communities.

\section{Related Work}
\subsection{Community Finding}
\label{sec:relcomm}
\noindent\textbf{Finding non-overlapping communities.} 
Algorithms in this context can be broadly grouped into three types. \textit{(1) Hierarchical algorithms} construct a tree of communities based on the network topology. These can be one of two types: divisive algorithms \cite{girvan2002community} or agglomerative algorithms \cite{clauset2004finding}. \textit{(2) Modularity-based algorithms} optimize the well-known modularity objective function to uncover communities in a network \cite{newman2006modularity}. \textit{(3) Other algorithms} which include those based on label propagation approaches \cite{xie11slpa}, spectral methods that make use of the eigenvectors of a graph's adjacency matrix, and methods based on statistical modeling \cite{fortunato10review}.
\vskip 0.7em
\noindent\textbf{Finding overlapping communities.} 
Existing algorithms in this context can be classified into four main categories. \textit{(1) Node seeding and local expansion algorithms} detect communities by starting from a node or a small group of nodes, then expanding them into a community using some fitness function. OSLOM \cite{lancichinetti11oslom} is an example of such an algorithm, which expands communities based on a fitness function measuring the statistical significance of communities with respect to random variations. 
\textit{(2) Clique expansion methods} use a group of fully-connected nodes, called a clique, as the starting point for building larger communities. Greedy Clique Expansion (GCE) \cite{lee10gce} is an example of this type of algorithm. \textit{(3) Link clustering algorithms} detect communities by splitting the network edges rather than the nodes \cite{amelio2014overlapping}. \textit{(4) Label propagation algorithms} attempt to group each node into a community based on its neighboring nodes' affinities. 
\vskip 0.7em
\noindent\textbf{Speaker-listener label propagation.} 
A representative example of this strategy is the Speaker Label Propagation Algorithm (SLPA) \cite{xie11slpa}. Here every node is associated with a corresponding \emph{memory} to store the frequencies of labels received from other nodes. Each node can take the role of either a \emph{listener} or a \emph{speaker}, and the roles are switched based on the state of the node -- \ie whether a node is providing information or consuming it. In the listener state, a node accepts labels from its neighbors, based on certain rules. In the speaker state, the node chooses a label from its own memory according to certain rules and sends it to neighboring listener nodes. Initially each node is assigned its own unique label. Then an iterative evaluation stage is repeatedly applied:
\begin{enumerate}
\item Randomly select one node as a listener.
\item Each neighbor of the listener randomly chooses a label from its own memory with a probability proportional to the frequency of occurrence of this label, and sends the label to the listener.
\item The listener chooses the most popular label among the received labels, and then adds it to its own memory.
\end{enumerate}
A subsequent post-processing stage converts each node's memory into a probability distribution of labels. If the probability of the frequency of a certain label is less than a user-specified threshold, the label is removed from a node's memory. After this thresholding step, all nodes having the same label are grouped into one community. Nodes that have more than one label naturally  belong to multiple communities
\subsection{Semi-Supervised Learning in Community Finding}
Several forms of prior knowledge have been used to guide community detection. The most widely-used strategy has been that of \emph{pairwise constraints} involving ``must-link'' and ``cannot-link'' relations. These relations indicate that either two nodes must be in the same community or must be in different communities. Such constraints have been implemented in several algorithms, including a modularity-based method \cite{li2014extremal}, a spectral analysis method \cite{habashi2016enhanced,zhang2013community}, and methods based on matrix factorization \cite{zhang2013community}. Instead of constraints, some authors have proposed the use of \textit{node labels} to encode prior knowledge for community detection \cite{leng2013active}. In \cite{liu2015effective}, the authors propose a method that uses a semi-supervised label propagation algorithm based on \textit{node labels} and \textit{negative information}, where a node is deemed to not belong to a specific community. 

The vast majority of semi-supervised algorithms in this area aim solely at detecting disjoint communities, whereas many real-world social networks contain overlapping structures \cite{ahn10link}. In \cite{dreier2014overlapping}, a small set of nodes called \textit{seed nodes} was used, whose affinities to a community is provided as prior knowledge to infer the rest of the nodes affinities in the network. However, to the best of our knowledge, no work has been done in the context of finding overlapping communities using supervision encoded as pairwise constraints. 


\section{Methods}
\subsection{Pairwise Constraints for Overlapping Communities}
Before describing the proposed methods for semi-supervised community finding, we firstly discuss the issue of selecting appropriate pairwise constraints for networks containing overlapping communities.

Given a network that contains a set of nodes $V$, semi-supervised pairwise constraints typically take two possible forms:
\begin{enumerate}
  \item A \textit{must-link constraint} specifies that two nodes should be in the same community. Let $C_{ML}$  be the must-link constraint set: $\forall$ $v_{i}, v_{j}$ $\in$ $V$ where  $i$ $\neq$ $j$, ($v_{i}, v_{j}$) $\in$ $C_{ML}$  indicates that two nodes $v_{i}$ and $v_{j}$ must be assigned to the same community. 
  \item A \textit{cannot-link constraint} specifies that two nodes should be in different communities. Let $C_{CL}$  be the cannot-link constraint set: $\forall$ $v_{i}, v_{j}$ $\in$ $V$ where  $i$ $\neq$ $j$, ($v_{i}, v_{j}$) $\in$ $C_{CL}$  indicates that $v_{i}$ and $v_{j}$ must be assigned to separate communities. 
\end{enumerate} 

These constraints are provided by the oracle, typically an individual expert or committee of annotators. 
The simplest approach for selecting pairwise constraints to present to the oracle is to na\"{\i}vely select a pair of nodes ($v_{i}, v_{j}$) at random, and query the oracle about whether the pair share a must-link or cannot-link relationship. This process is typically repeated until some supervision budget is exhausted.

\begin{figure}[!t]
    \centering
        \subfloat[Non-overlapping case]{\includegraphics[width=0.31\columnwidth]{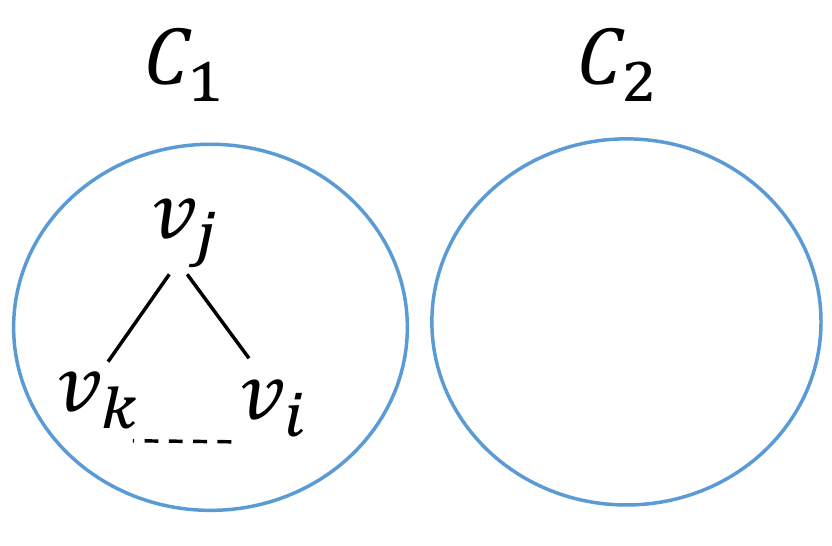}}\hspace{2.1em}
        \subfloat[Overlapping case]{\includegraphics[width=0.58\columnwidth]{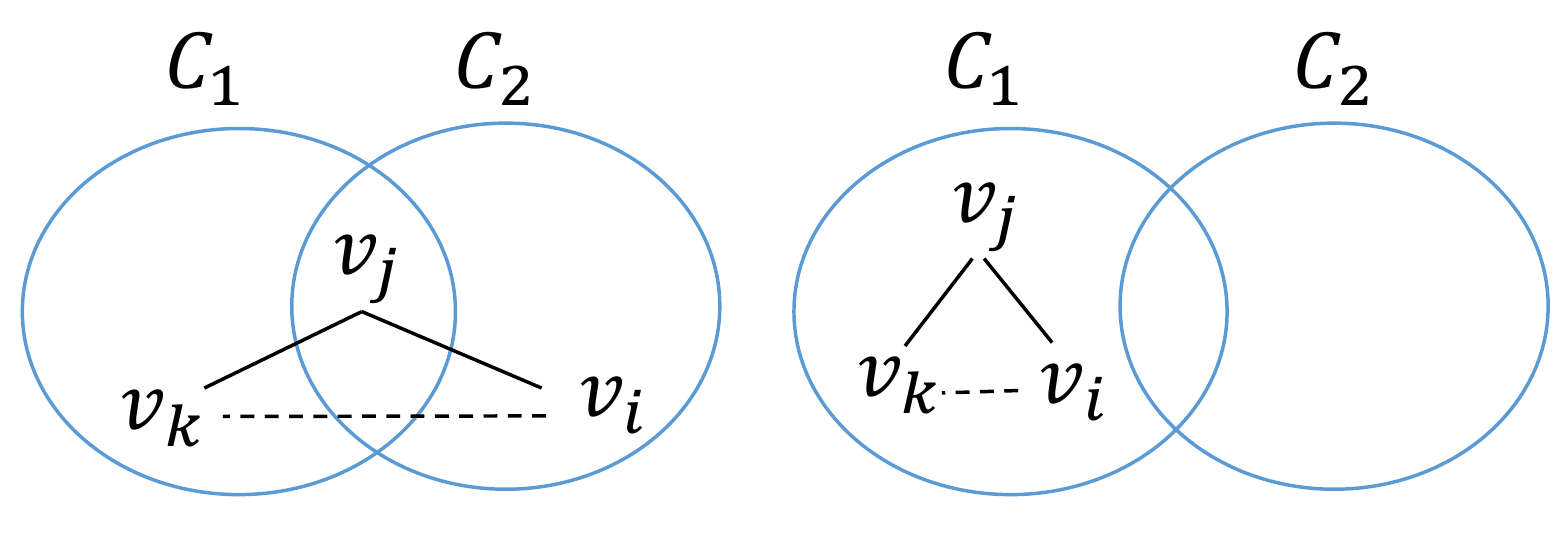}}
    \caption{In the non-overlapping case (a), the transitive property allows us to infer a third must-link constraint from two existing must-link constraints. However, this does not automatically apply in the overlapping case (b), where two possible situations exist.}
	\label{fig:transitive}
\end{figure}

In non-overlapping community finding, must-link constraints have a \textit{transitive property}, such that a third must-link relationship can be inferred from two other associated must-link constraint pairs. So, if ($v_{i}, v_{j}$) $\in$ $C_{ML}$, and ($v_{i}, v_{k}$) $\in$ $C_{ML}$, then we can also infer that ($v_{j}, v_{k}$) $\in$ $C_{ML}$ (see Fig. \ref{fig:transitive}(a)). 

However, incorporating constraints into the context of overlapping communities is more challenging. This is because the transitive property does not hold here (see the second example in Fig. \ref{fig:transitive}). Specifically, if ($v_{i}, v_{j}$) $\in$ $C_{ML}$, and ($v_{i}, v_{k}$) $\in$ $C_{ML}$, there are two possible scenarios for the pair ($v_{j}, v_{k}$). It can be the case that either ($v_{j}, v_{k}$) $\in$ $C_{ML}$ or ($v_{j}, v_{k}$) $\in$ $C_{CL}$. This is because an overlapping node $v_{j}$ can have a must-link constraint with both $v_{i}$ and $v_{i}$, yet these two nodes could belong to two different communities. However, it is also possible that all three nodes are in fact in the same community. Unless we explicitly inform the algorithm about whether a must-link or cannot-link constraint exists for the pair ($v_{j}, v_{k}$), the algorithm cannot reliably distinguish between the two cases.

If the network has highly-overlapping communities (\ie each node typically belongs to many communities), then this problematic situation will occur more frequently. Therefore, if we attempt to incorporate pairwise constraints into overlapping community finding without taking this situation into account, the quality of the resulting communities can potentially decrease, even as more constraints are added. Next we introduce a strategy to resolve this issue.

\subsection{Semi-Supervised Overlapping Community Finding}
We now propose a new semi-supervised label propagation procedure for finding overlapping communities, which consists of two distinct phases: 
\begin{enumerate}
\item Select and pre-process constraints, to resolve the problem of the lack of the transitive property for must-link constraints. 
\item Apply label propagation-based community finding, in a manner that takes into account information provided by the selected constraints.
\end{enumerate}

\begin{figure}[!t]
\centering
\includegraphics[width=\linewidth]{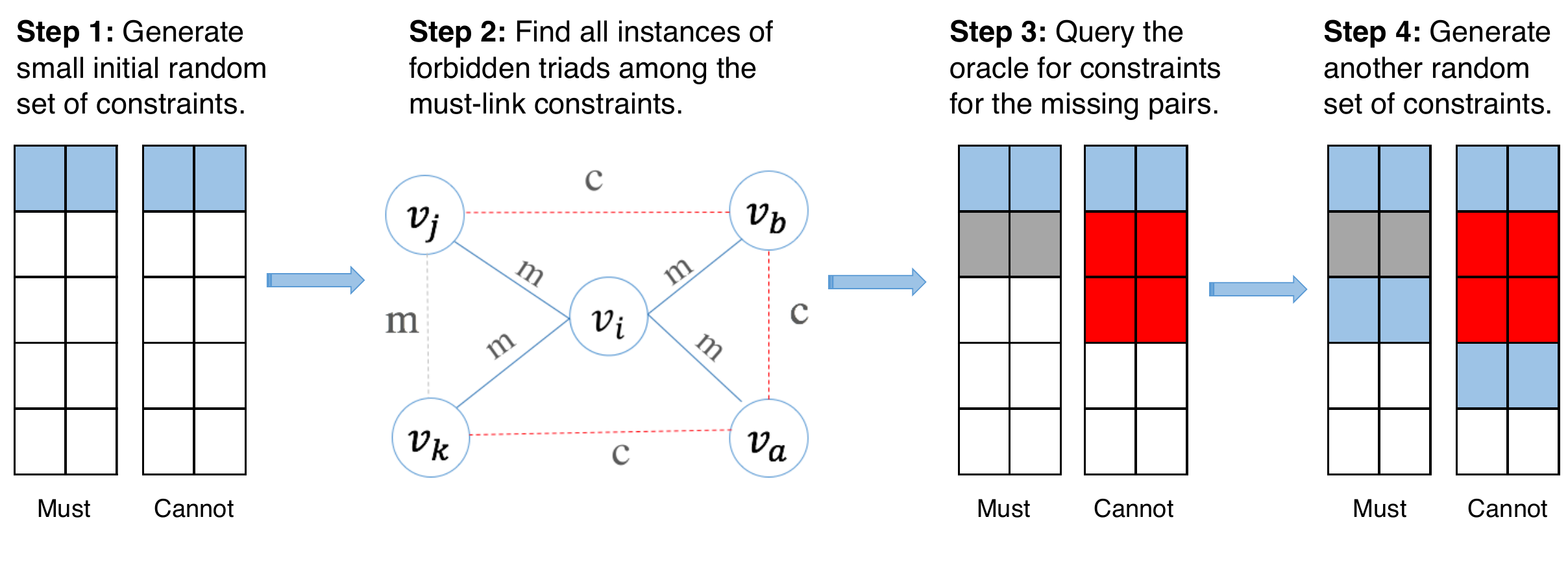}
  \caption{An illustration of all steps in the overlapping constraint selection process.}
  \label{fig:select}
\end{figure}

\vskip 0.3em
\noindent\textbf{Phase 1: Selecting and pre-processing constraints.} After selecting an initial set of pairwise constraints by querying an oracle, we can view the set of pairwise constraints as a new graph, where an edge exists between two nodes from the original network if they share a pairwise constraint (either must-link or cannot-link). Then we look for all possible \textit{forbidden triads} among the nodes involved in the must-link set. Given three nodes A, B, C, a \textit{forbidden triad} (sometimes referred to as an \textit{open triad}) occurs when A is connected to B and C, but no edge exists between B and C. In our pre-processing step, we look for such cases --- \ie where we do not know whether a must-link or cannot-link exists between a pair of nodes B and C. To control the size of the constraints set, we greedily expand it until we reach a pre-defined maximum size. 
The complete constraint selection strategy can be summarized as follows (see also Fig. \ref{fig:select}):
\begin{enumerate}
  \item Select a small random set of both must-link and cannot-link constraints.
  \item Find all possible forbidden triads in the must-link set, to identify pairs to query the oracle about their relationship.
  \item For each resulting pair, if their relationship is must-link, then add the pair to the must-link set. Otherwise, add the pair to the cannot-link set.
  \item Repeat all steps until the maximum number of selected constraints is reached.
\end{enumerate}
At the end of this process, the pairwise constraints are ready to be supplied to the community detection algorithm, which we describe next.

\begin{figure}[!t]
\centering
\includegraphics[width=0.95\linewidth]{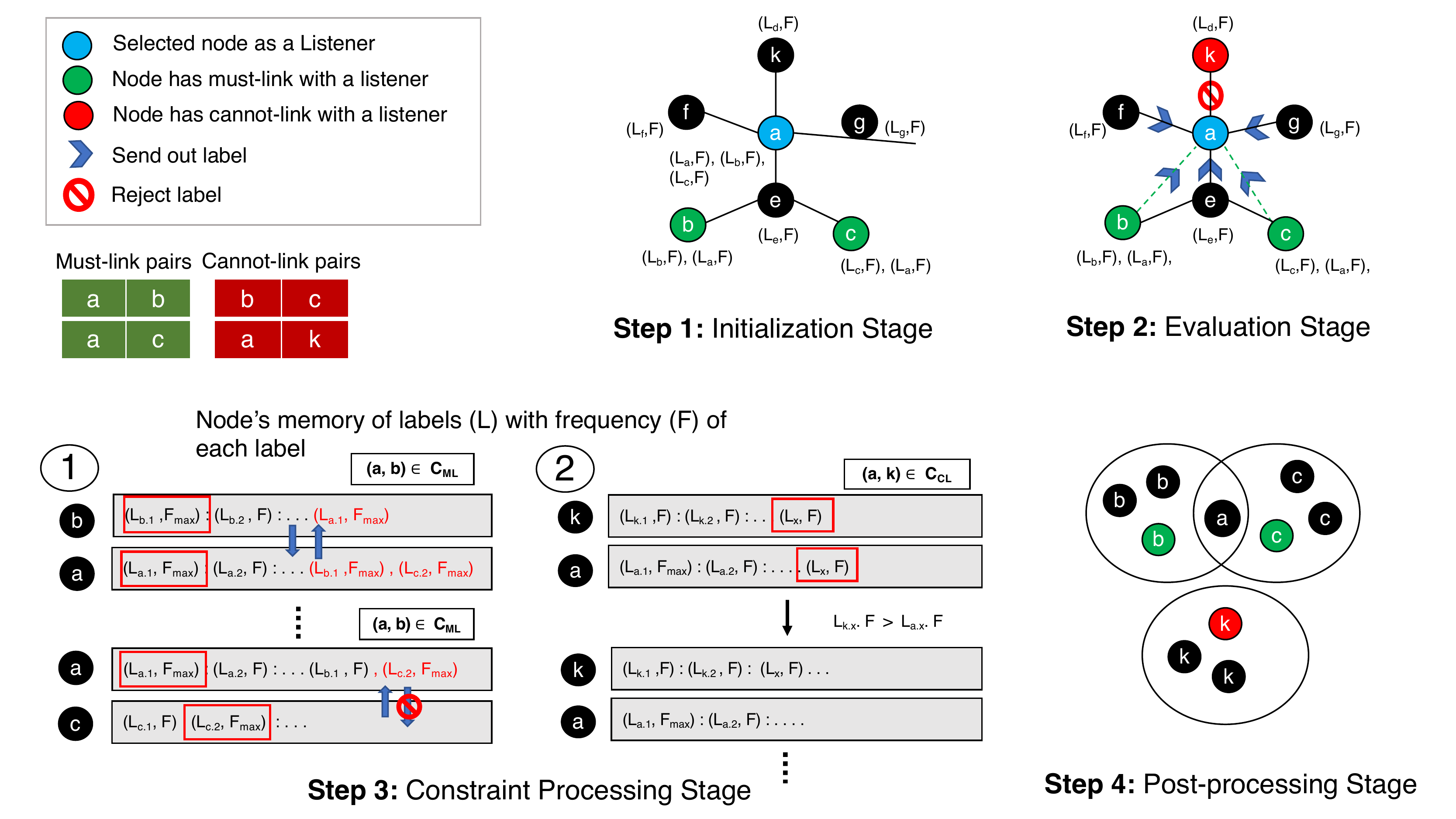}
\caption{An illustration of Steps 1-4 involved in the PC-SLPA algorithm.}
\label{fig:pcslpa}
\end{figure}

\vskip 0.7em
\noindent\textbf{Phase 2: Pairwise Constrained SLPA (PC-SLPA).} We incorporate the selected pairwise constraints as follows (see also Fig. \ref{fig:pcslpa}):
\begin{enumerate}
  \item In the \emph{initialization step}:
   \begin{enumerate}
      \item Give a unique label to each node in the network. 
      \item For each pair of nodes having a must-link relationship, the two nodes exchange labels (\ie update each node's memory with the other node's label). 
  \end{enumerate}
\vskip 0.5em
\item  The \emph{evaluation step} broadly follows a similar process as unsupervised SLPA (see Section \ref{sec:relcomm}). However, we account for the pairwise constraints as follows:
\begin{enumerate}
\item Randomly select one node as a listener, and identify the set of speakers (\ie the neighbours of the listener).
  \item Augment the set of speakers by adding all nodes that have must-link relationship with the listener and removing all nodes that have cannot link relationship with the listener. Then each speaker sends out a label according to the rule defined in standard SLPA.
\item Each listener accepts the sent labels, unless sent by a node which shares a cannot-link constraint with the listener. Since all nodes that hold the same label will be grouped together as a community at the end of the process, this avoids grouping together pairs of nodes having a cannot-link relationship.
\end{enumerate}
\vskip 0.5em
\item The \emph{constraint processing step} considers both sets of pairwise constraints:
\begin{enumerate}
\item For each must-link pair, compare the memories of the two nodes to ensure they both share the same highest occurrence frequency label. If they do not, both nodes exchange their most frequently-occurring labels with each other under a condition that each node does not has a cannot link relationship with any node assign to that label. 
\item For each cannot-link pair, compare the memories of both nodes. If both nodes have a common label, remove this label from the node that has the lowest label occurrence frequency. 
\end{enumerate}
\vskip 0.5em
\item In the \textit{post-processing step}, convert each node's memory into a probability distribution of labels. If the node's probability of a certain label is less that a threshold $r \in [0, 1]$, the label is removed from the node's memory. Then all nodes having the same label are grouped into one community. Nodes that have more than one label correspond to overlapping cases which belong to multiple communities.
\end{enumerate}



\section{Evaluation}
\label{sec:eval}

\subsection{Experimental Setup}
\label{sec:data}

We now evaluate the performance of PC-SLPA to determine the extent to which introducing varying levels of constraints can improve community detection. 

\vskip 0.3em
\noindent\textbf{Data.} Firstly, we evaluate on synthetic data created using the widely-used LFR generator \cite{lancichinetti2008benchmark}, which can produce networks with properties similar to real-world networks, with overlapping ground truth communities. The selection of network parameters shown in Table 1(a) is based on those used to evaluate the original algorithm SLPA \cite{xie11slpa} and other works in the literature. We generate two different groups of synthetic networks with different sizes, each containing small and large communities and mixing parameter $\mu$ varies from 0.1 to 0.3. Small communities have 10--50 nodes, while large communities have 20--100 nodes. Each group consists of 16 networks with different combinations of the parameter $O_m$, which controls the number of communities per node. For the first network in each set, all nodes belong to two communities ($O_m=2$). For each successive network, this parameter value is incremented by 1 until $O_m=8$ is reached. 

Secondly, we consider three real-world networks which have previously been used in the community finding literature \cite{leskovec2015snap}: 1) a co-purchasing network from Amazon.com; 2) a friendship network from YouTube; 3) a scientific collaboration network from DBLP.  These networks contain annotated ground truth overlapping communities. For each network, we include only the 5,000 largest such communities, as per \cite{yang2015defining}. We then perform filtering as per \cite{harenberg2014community} -- the remaining communities are ranked based on their internal densities and the bottom quartile is discarded, along with any duplicate communities. 
Finally, as an additional step, we eliminated extremely small communities. For the Amazon and YouTube networks, communities of size $<5$ nodes are removed, while for the DBLP network communities with $<10$ nodes are removed. Details of the resulting networks are listed in Table 1(b). 

\vskip 0.3em
\noindent\textbf{Baselines.} To the best of our knowledge, no work has been conducted in the literature regarding pairwise constrained algorithms for finding overlapping communities. Therefore, for the sake of comparison, the PC-SLPA results are compared with outputs of the following popular unsupervised overlapping community detection algorithms: SLPA\cite{xie11slpa}, OSLOM \cite{lancichinetti11oslom}, MOSES \cite{mcdaid2010detecting}, and COPRA \cite{gregory2010finding}. For OSLOM and MOSES, we use the default parameters recommended by the original authors. For COPRA, we use the settings recommended in \cite{xie11slpa}. To evaluate the performance of these algorithms relative to the ground truth groupings, we use the overlapping form of Normalized Mutual Information (NMI) \cite{lancichinetti2009detecting}. Since SLPA and COPRA are non-deterministic, we average the NMI values over 20 runs. 

\vskip 0.3em
\noindent\textbf{Experiments.} We conducted two experiments in our evaluation. The first aims to assess the performance of the unsupervised algorithms, which provides a baseline for evaluating the performance of our proposed method. For both SLPA and PC-SLPA we use the default parameters values $T=100$ and $r \in [0, 1],$ as suggested in \cite{lee10gce}. The second experiment evaluates the performance of PC-SLPA with increasing numbers of constraints, from $1\%$ to $5\%$ of the total number of possible pairs in each network. Since the initial pairwise constraints are selected at random, we repeat the semi-supervised process for 20 runs and average the resulting NMI scores.

\begin{table}[!b]
\centering
\caption{The first table lists parameters used for the generation of LFR synthetic networks. The second table summarizes details of the real-world networks.}
\vskip 0.5em
\scriptsize{
 \begin{tabular}{|c|l|c?c|l|c|}
\thickhline
Parameter & \emph{Description} & \emph{Value} & Parameter & \emph{Description} & \emph{Value} \\ [2pt]  \thickhline
$N$ & Number of nodes & 1000-5000 & $t_1$ & Degree exponent & 2 \\ \hline
$k$ & Average degree & 10 & $t_2$ & Community exponent & 1 \\ \hline
$K_{max}$ & Max degree & 50 & $\mu$ & Mixing parameter & 0.1-0.3 \\ \hline
$C_{min}$ & Min community size\; & 10/20 & $O_m$ & Communities per node\; & 1-8 \\ \hline
$C_{max}$ & Max community size\; & 50/100 &  &  & \\ \thickhline
\end{tabular}}
\vskip 0.8em
\scriptsize{
\begin{tabular}{|l?l|l|l|}
\thickhline
Real-world Networks & \emph{Amazon} & \emph{YouTube} & \emph{DBLP} \\ \hline
\emph{\#Nodes - \# Edges - \#Communities}\; & 7411 - 21214 - 876 & 6426 - 23226 - 1058 & 7233 - 33045 - 613 \\ \hline
\emph{Max community size} & 27 & 31 & 38 \\ \hline
\emph{Min community size} & 5 & 5 & 10 \\ \hline
\emph{Max communities per node} & 4 & 11 & 8\\ \hline
\emph{\#Overlapping nodes} & 1394(18\%) & 865(13\%) & 214 (3.3\%)\\  \thickhline
\end{tabular}}
\end{table}

\subsection{Results and Discussion}
\begin{figure}[!t]
\centering
\subfloat{\includegraphics[width=0.44\linewidth,trim={0.3cm 0.1cm 0.1cm 0},clip]{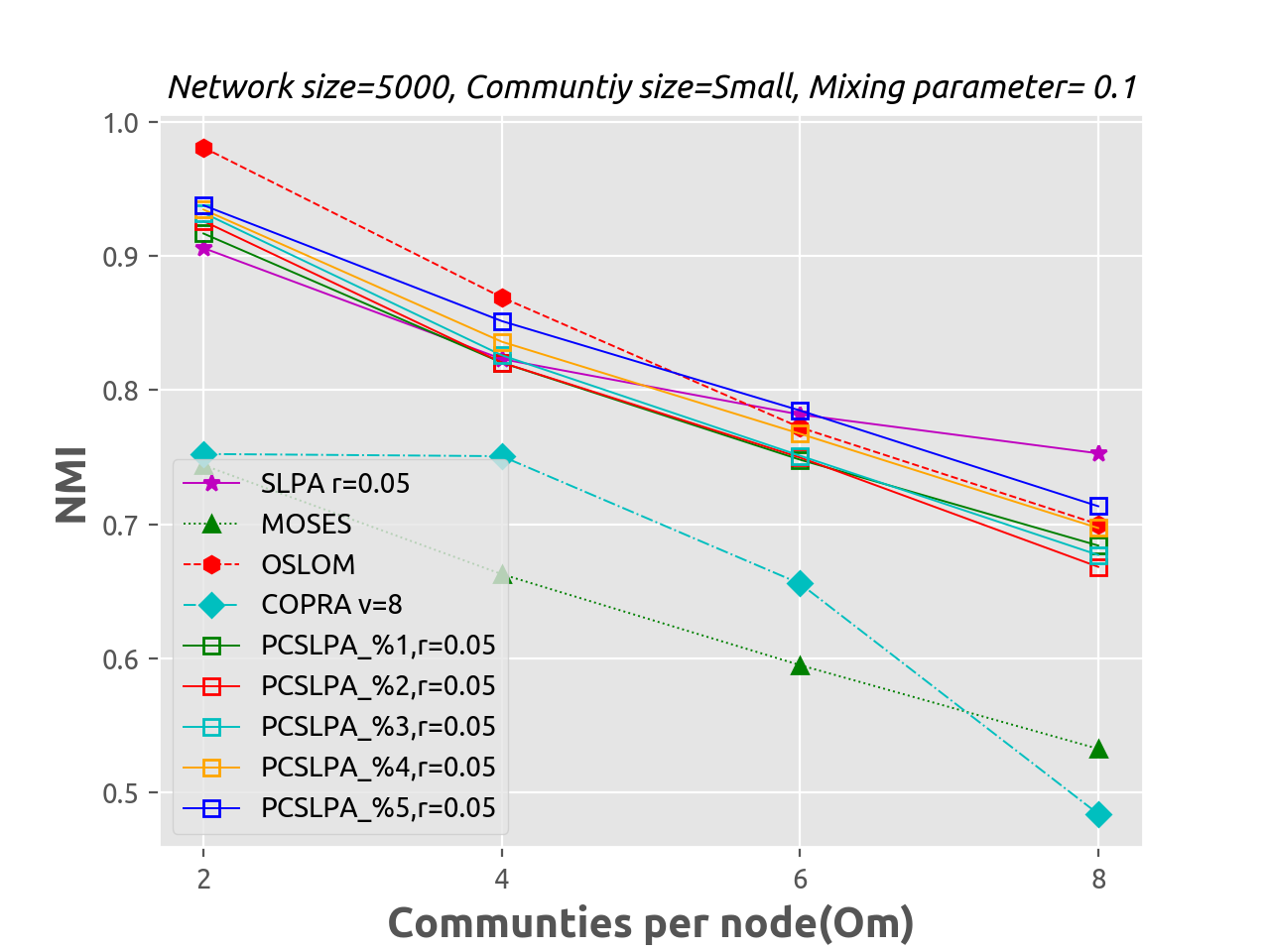}}
\subfloat{\includegraphics[width=0.44\linewidth,trim={0.3cm 0.1cm 0.1cm 0},clip]{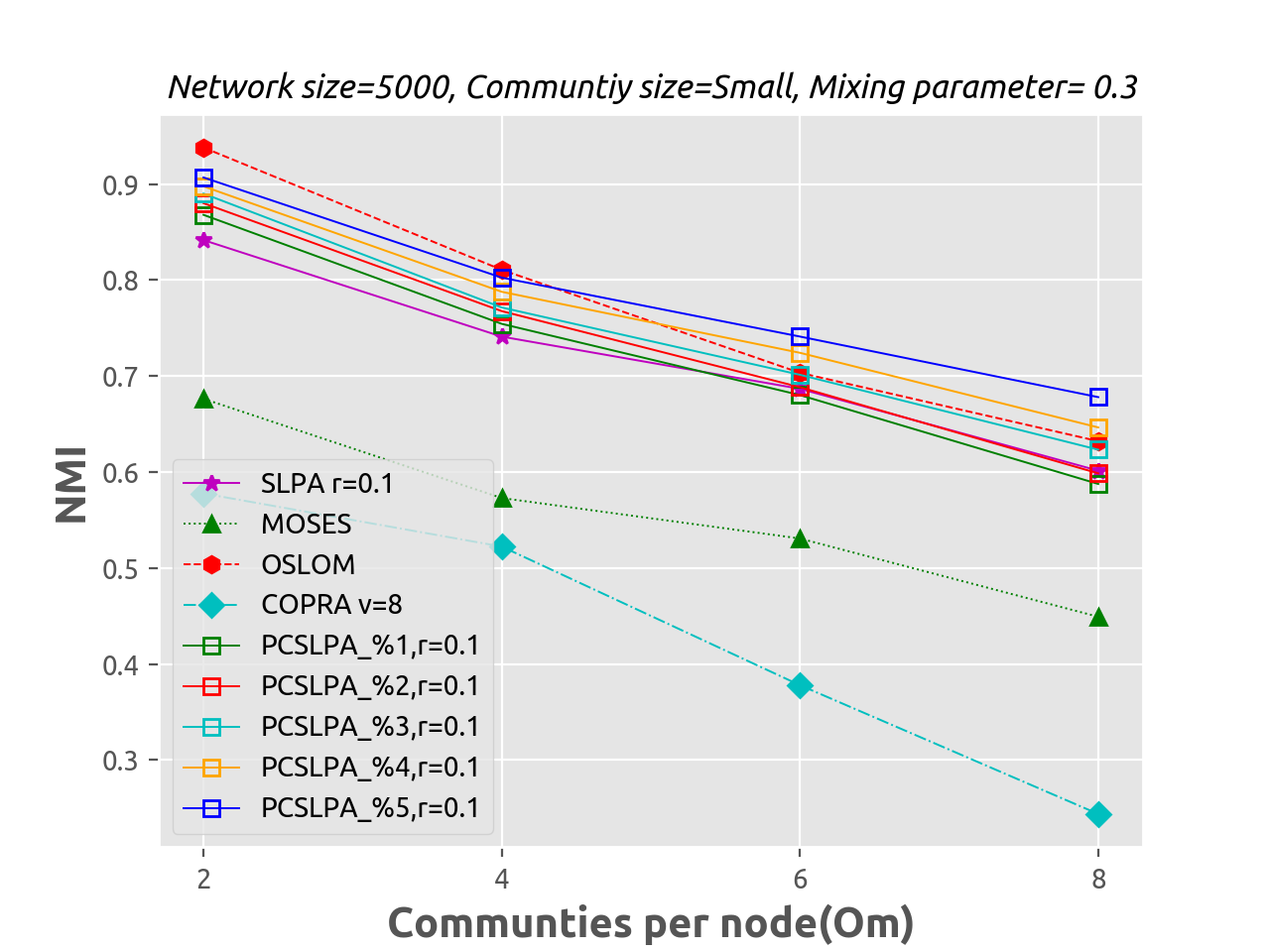}}
\vskip -0.9em
\subfloat{\includegraphics[width=0.44\linewidth,trim={0.3cm 0.1cm 0.1cm 0},clip]{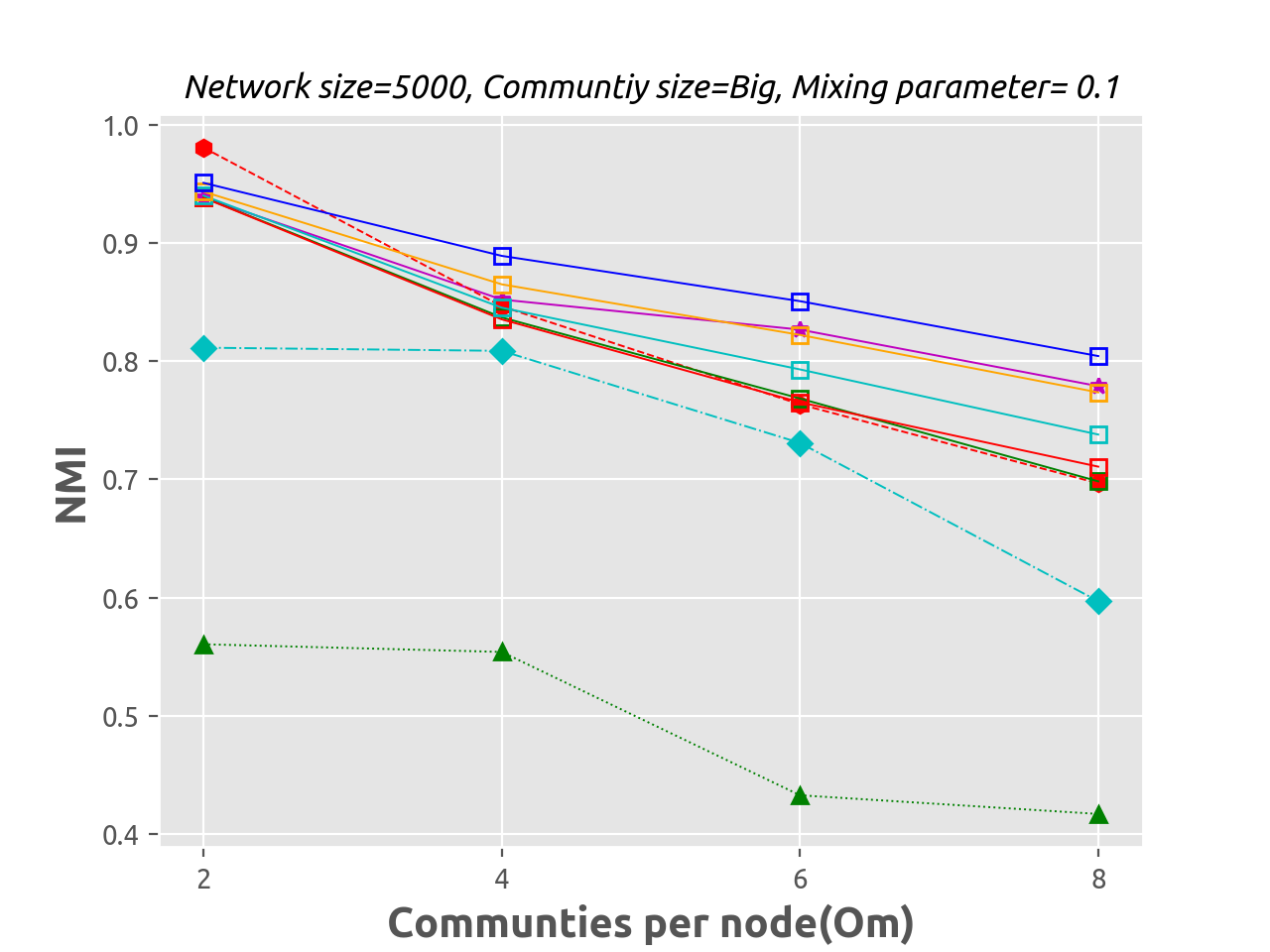}}
\subfloat{\includegraphics[width=0.44\linewidth,trim={0.3cm 0.1cm 0.1cm 0},clip]{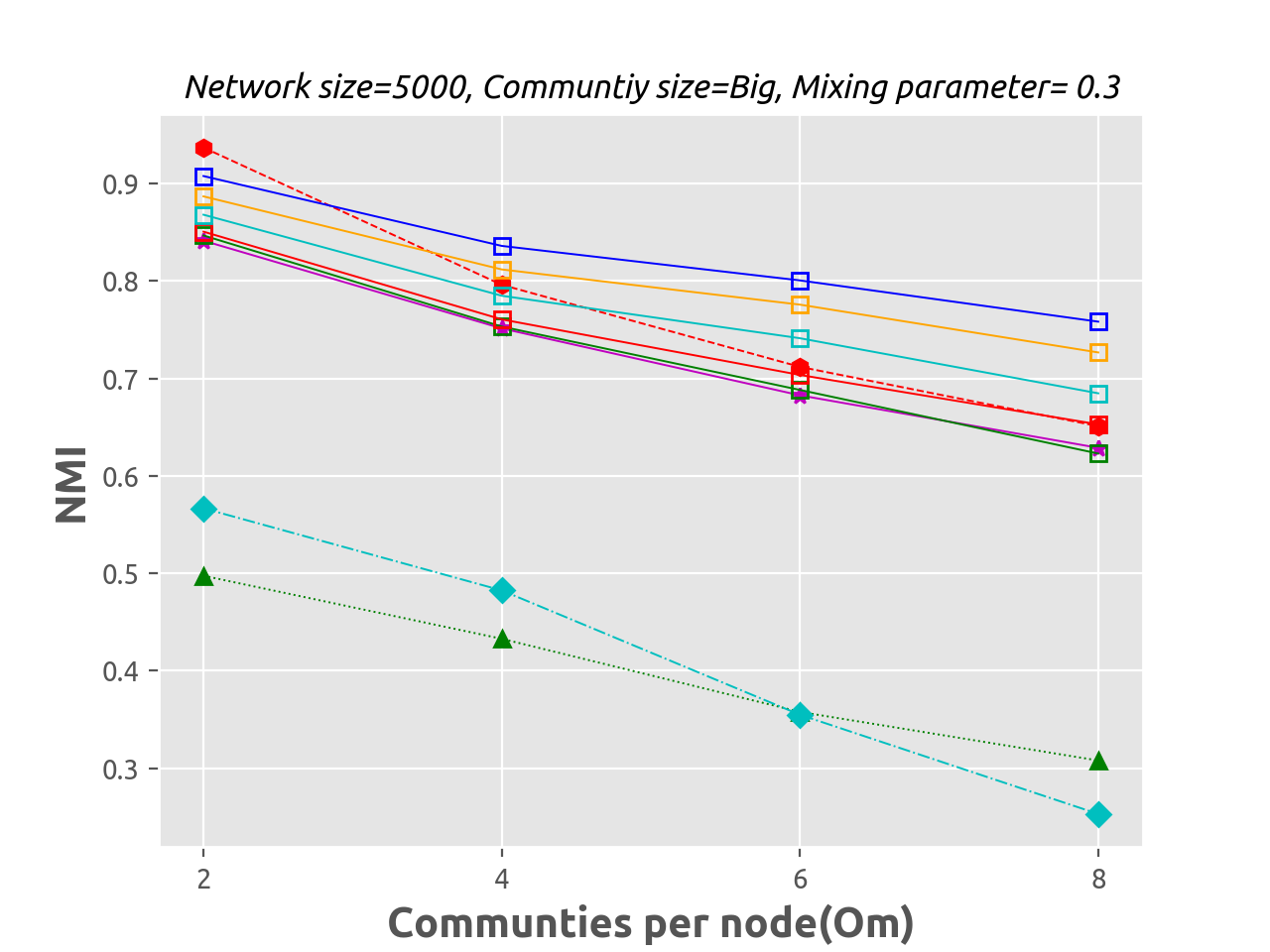}}
\vskip -1.1em
\subfloat{\includegraphics[width=0.44\linewidth,trim={0.3cm 0.1cm 0.1cm 0},clip]{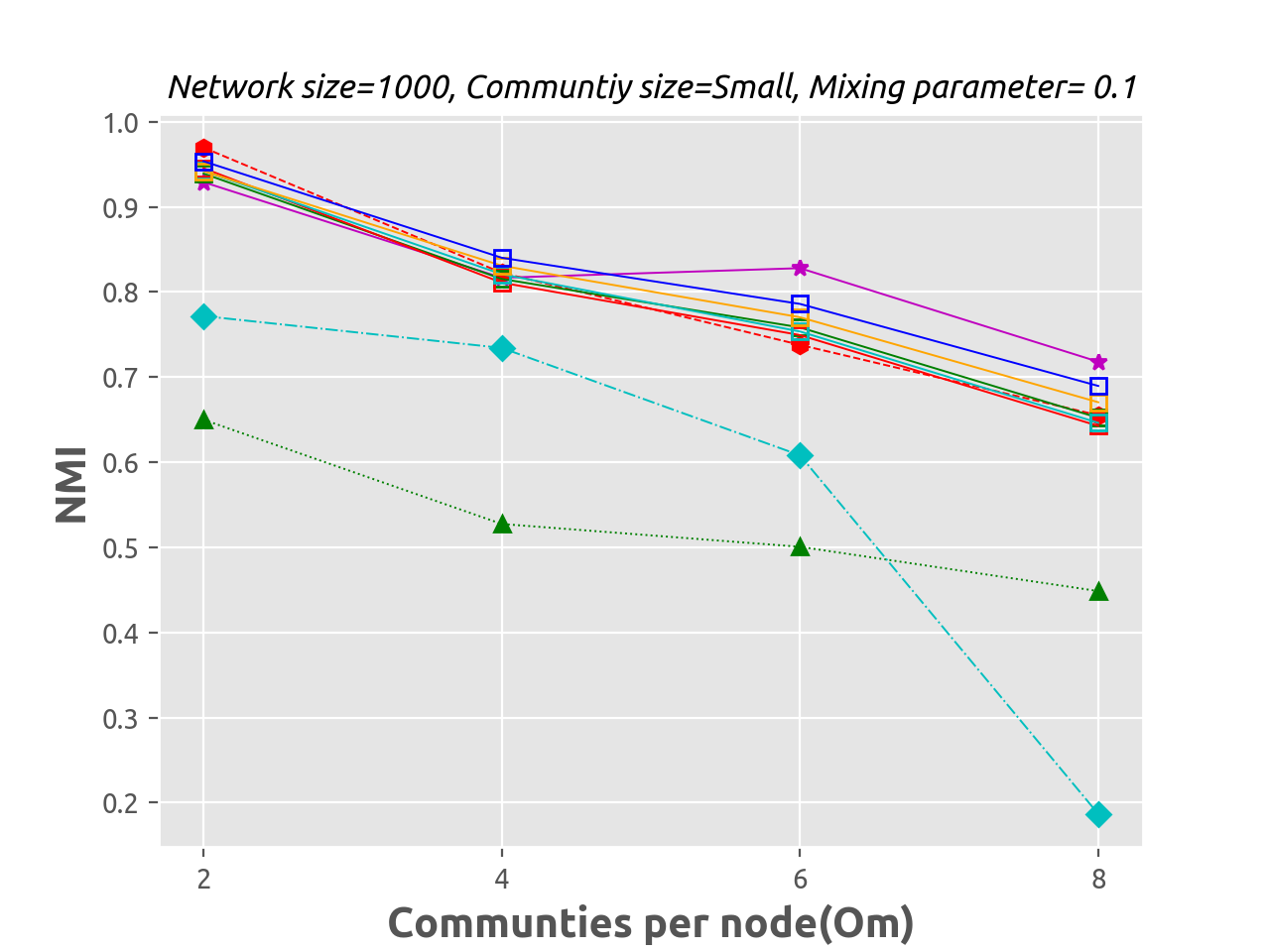}}
\subfloat{\includegraphics[width=0.44\linewidth,trim={0.3cm 0.1cm 0.1cm 0},clip]{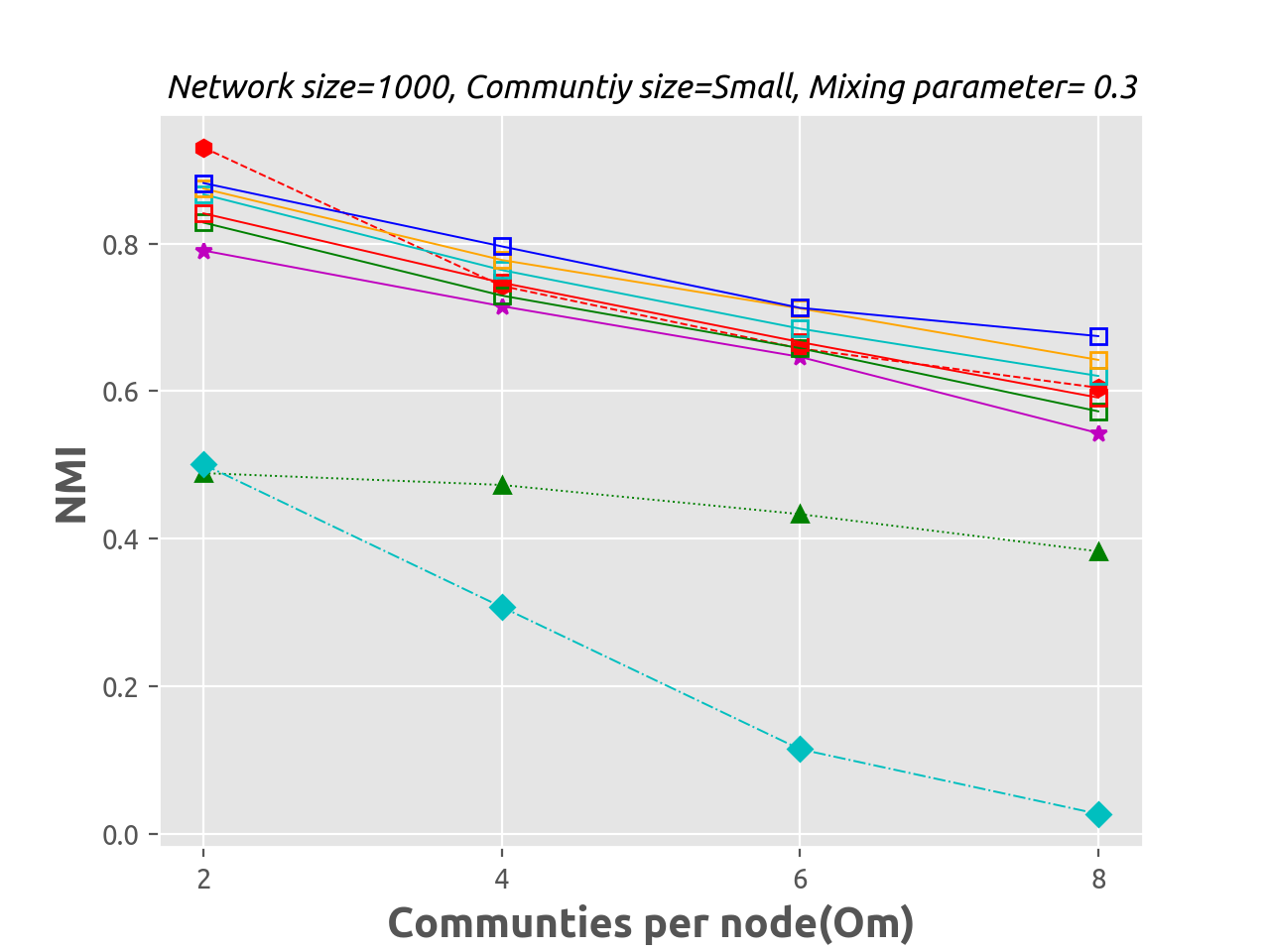}}
\vskip -1.1em
\subfloat{\includegraphics[width=0.44\linewidth,trim={0.3cm 0.1cm 0.1cm 0},clip]{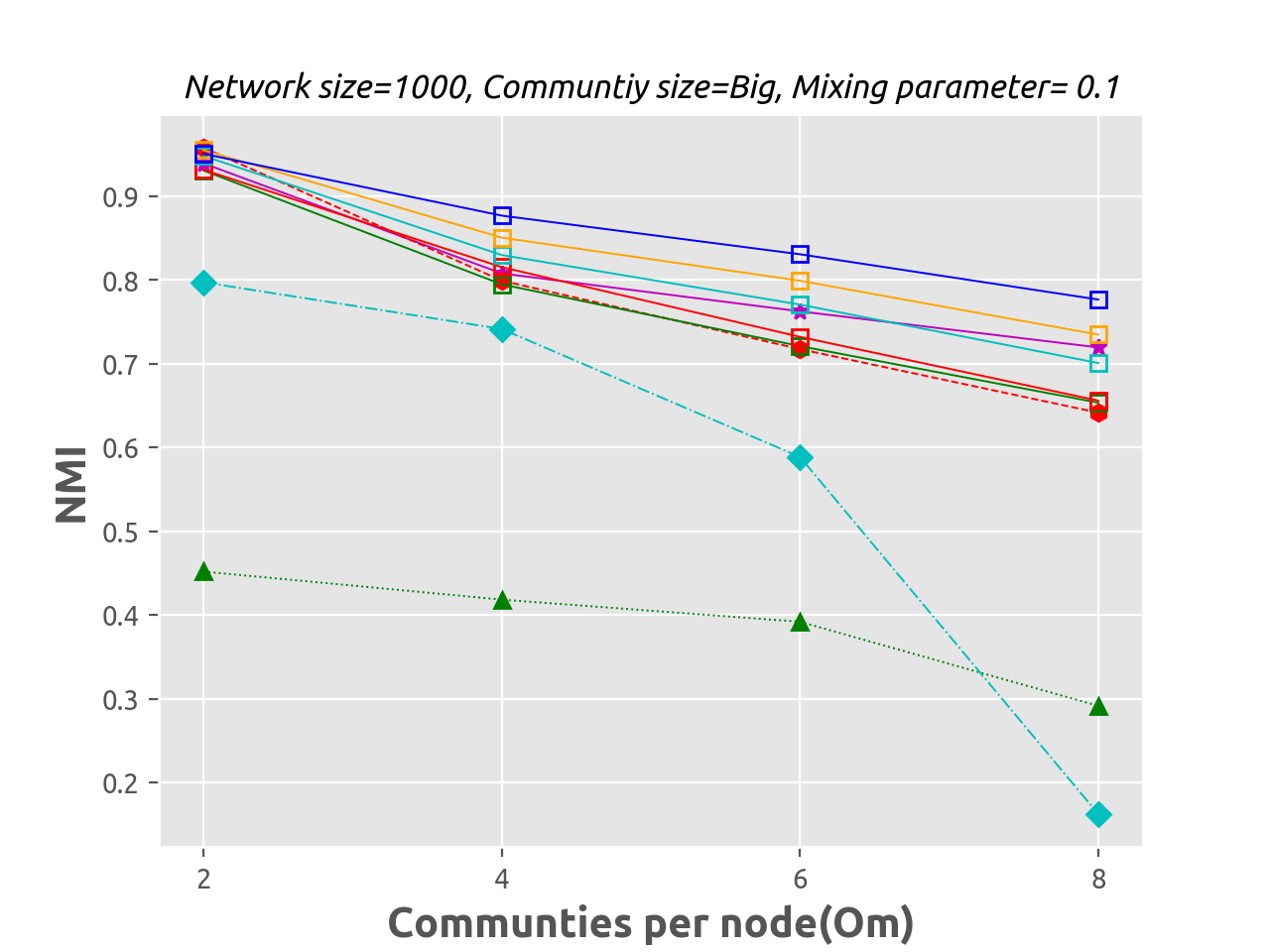}}
\subfloat{\includegraphics[width=0.44\linewidth,trim={0.3cm 0.1cm 0.1cm 0},clip]{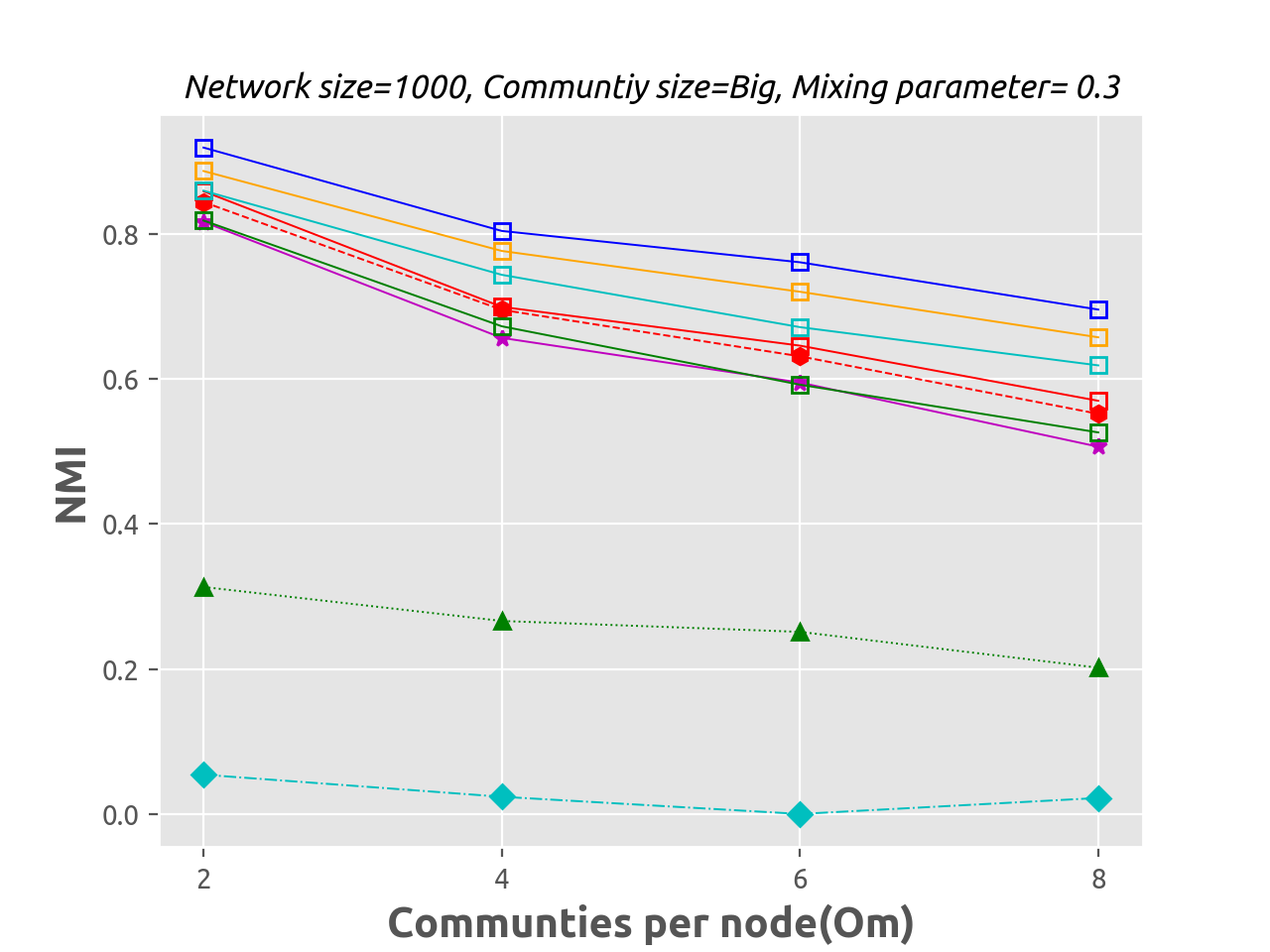}}
\vskip 0.3em
\caption{Performance of all algorithms on 32 synthetic networks, containing both small and large communities, with mixing parameter $\mu \in [0.1,0.3]$. NMI values are plotted against the number of communities per node $(O_m)$, with 4 networks in each plot.}
\label{NMIRE}
\end{figure}

\noindent\textbf{Synthetic networks.} For most of the 32 networks, PC-SLPA achieves consistently higher NMI scores than the standard SLPA algorithm, except where $\mu = 0.1$. Here PC-SLPA attains lower NMI values than SLPA, until the number of constraints increases towards $5\%$. In general, as the percentage of pairwise constraints being used increases, the accuracy of PC-SLPA improves significantly.

When evaluating on LFR-generated networks, different factors can affect algorithm performance, such as the mixing parameters, and the size of both networks and embedded communities. The larger the value of $\mu$, the poorer the communities detected algorithms due to the weaker intra-community connectivity. As we see from Fig.\ref{NMIRE}, the performance of SLPA drops as $\mu$ increases from 0.1 to 0.3. However, PC-SLPA shows more stability with higher values of $\mu$. For instance, in the case of small networks of big communities with $\mu=0.3$, the NMI score of the standard SLPA is 0.82 at $O_m=2$ and drops to 0.50 when $O_m$ increased to 8. In contrast, the PC-SPLA algorithm shows a more moderate decrease in accuracy as the value of $O_m$ increases. As for the size of network, both algorithms show better performance when the network  increases from 1,000 to 5,000 nodes, with PC-SLPA achieving the best performance on the networks with larger communities.

When comparing PC-SLPA to the baseline algorithms, we observe that COPRA and MOSES show the lowest performance on all synthetic networks. As for OSLOM, it shows slightly better performance than PC-SLPA for networks with a low level of community overlap. However, as the number of communities per node increases, PC-SLPA starts to out-perform all of the baseline algorithms, indicating that it is effective in highly-overlapping contexts.

Table \ref{table3} summarizes the performance of all algorithms as win-loss records. Each table entry shows the number of wins of an algorithm (on the rows) over another algorithm (on the columns). To compare two algorithms, we subtract the sum of wins and losses from the total number of synthetic networks. The last column reports rank scores based on the total number of ``wins'' by each algorithm across all synthetic networks. According to the total number of wins, we rank the highest number as the best algorithm. Then, we order the algorithms from best to worst. As we can see from Table \ref{table3}, PC-SLPA with $5\%$ pairwise constraints is the top-ranked algorithm, and performs better than the competing benchmark algorithms on $90\%$ of the networks. OSLOM is the next best alternative, followed by SLPA. 

\begin{table}[!b]
\centering
\caption{Win-loss table of NMI performance for all algorithms on 32 synthetic networks.}
\vskip 0.3em
\label{table3}
\scriptsize{
\begin{tabular}{|l|l|l|l|l|l|l?l|l|}
\thickhline
\multicolumn{2}{|l|}{\multirow{2}{*}{}} & \multicolumn{5}{l?}{\emph{Loser}} & \multicolumn{2}{l|}{\emph{Rank-score}} \\ \cline{3-9} 
\multicolumn{2}{|l|}{} & \emph{OSLOM}\; & \emph{MOSES}\; & \emph{COPRA}\; & \emph{SLPA}\; & \emph{PCSLPA5\%}\; & \emph{Total wins}\; & \emph{Rank}\; \\ \thickhline
\multirow{5}{*}{\emph{Winner}} 
& \emph{PCSLPA5\%}\; & 22 & 32 & 32 & 29 & 0 & 115/128 (90\%) & 1 \\ \cline{2-9}
& \emph{OSLOM} & 0 & 32 & 32 & 22 & 10 & 96/128 (75\%) & 2 \\ \cline{2-9}
& \emph{SLPA} & 10 & 32 & 32 & 0 & 3 & 77/128 (60\%) & 3 \\ \cline{2-9} 
& \emph{MOSES} & 0 & 0 & 16 & 0 & 0 & 22/128 (12.5\%) & 4 \\ \cline{2-9} 
 & \emph{COPRA} & 0 & 16 & 0 & 0 & 0 & 10/128 (12.5\%) & 4 \\ \thickhline
\end{tabular}}
\end{table}

\vskip 0.4em
\noindent\textbf{Real-world networks.} Next we discuss our experiments on the three real-world networks. We compare the NMI performance of our proposed semi-supervised method with increasing numbers of pairwise constraints, relative to the benchmark algorithms. For the non-deterministic algorithms, 20 runs were executed and NMI scores were averaged.
From Table \ref{table2}, we see that PC-SLPA algorithm achieves high NMI scores $(> 0.9)$ on the Amazon and DBLP networks. However, PC-SLPA shows moderate performance on YouTube network, which may be due to the poor separation between the ground truth groups in this network. The addition of $<4\%$ of constraints does not yield an improvement over the unsupervised approach. The effect of high inter-community overlap is far more pronounced in the cases of the OSLOM, MOSES, and COPRA algorithms. Overall, PC-SLPA outperforms the four alternative algorithms in most cases on these networks, with small but consistent increases as the number of provided constraints is increased from 1\% to 5\%. We would expect this trend to continue as more constraints are added, although it may be impractical to generate larger numbers of constraints in real-world scenarios.

\begin{table}[!t]
\centering
\caption{NMI scores of all algorithms on three real-world networks.}
\vskip 0.3em
\label{table2}
\resizebox{\columnwidth}{!}{
\begin{tabular}{|l?c|c|c|c?c|c|c|c|c|}
\thickhline
 & \;\emph{OSLOM}\; & \;\emph{MOSES}\; & \;\emph{COPRA}\; & \;\emph{SLPA}\; & \emph{PCSLPA\%1} & \emph{PCSLPA\%2} & \emph{PCSLPA\%3} & \emph{PCSLPA\%4} & \emph{PCSLPA\%5} \\ \thickhline
Amazon & 0.9668 & 0.9084 & 0.96228 & 0.9568 & 0.9612 & 0.9650 & 0.9678 & 0.9709 & 0.9723 \\ \hline
YouTube \;\; & 0.4490 & 0.4209 & 0.1907 & 0.6296 & 0.6011 & 0.6130 & 0.6241 & 0.6338 & 0.6439 \\ \hline
DBLP & 0.8485 & 0.7707 & 0.9136 & 0.8972 & 0.9059 & 0.9156 & 0.9231 & 0.9278 & 0.9326 \\ \thickhline
\end{tabular}}
\end{table}



\section{Conclusion}

We have proposed a new algorithm, PC-SLPA, for detecting overlapping communities, based on the use of a label propagation strategy that is informed by the addition of external information encoded as pairwise constraints. We explored the nuances around the selection of constraints, which are specific to contexts where the communities in the data naturally overlap. Based on extensive experiments, the results show that overlapping community finding algorithms with constraints can considerably out-perform their unconstrained counterparts on both synthetic and real-world networks. As one might expect, their performance improves with increasing number of pairwise constraints. In general, the results show the potential of using semi-supervised strategies for finding overlapping communities. In our future work we will aim to apply ideas from active learning for selecting informative pairwise constraints, in order to reduce the annotation burden on the oracle, while maintaining or even improving the effectiveness of community detection.

\vskip 1.5em
\noindent \textbf{Acknowledgements.} This research was partly supported by Science Foundation Ireland (SFI) under Grant Number SFI/12/RC/2289.

%
%
\bibliographystyle{spmpsci}
\bibliography{bibliotest}

\end{document}